\documentclass[article]{jss}

\usepackage{orcidlink,thumbpdf,lmodern}
\usepackage{graphicx}
\usepackage{amsmath}
\usepackage{natbib}
\usepackage{longtable}
\usepackage{placeins}
\usepackage{tikz}
\usetikzlibrary{arrows.meta}
\tikzset{>={Latex[width=3mm,length=3mm]}}
\usetikzlibrary{shapes}

\usepackage{framed}

\newcommand{\bv}[1]{\boldsymbol{#1}}

\newcommand{\R}{\proglang{R} }

\usepackage{Sweave}

\author{Daniel K. Sewell~\orcidlink{0000-0002-9238-4026}\\University of Iowa
   \And Alan Arakkal~\orcidlink{0000-0002-7001-493X}\\JANE LLC}
\Plainauthor{Daniel K. Sewell, Alan T. Arakkal}

\title{bayesics: Core Statistical Methods via Bayesian Inference in \proglang{R}}
\Plaintitle{bayesics: Core Statistical Methods via Bayesian Inference in R}
\Shorttitle{bayesics: Core Statistical Methods via Bayesian Inference in \proglang{R}}

\Abstract{
  Bayesian statistics is an integral part of contemporary applied science.  \pkg{bayesics} provides a single framework, unified in syntax and output, for performing the most commonly used statistical procedures, ranging from one- and two-sample inference to general mediation analysis.  \pkg{bayesics} leans hard away from the requirement that users be familiar with sampling algorithms by using closed-form solutions whenever possible, and automatically selecting the number of posterior samples required for accurate inference when such solutions are not possible. \pkg{bayesics} focuses on providing key inferential quantities: point estimates, credible intervals, probability of direction, region of practical equivalance (ROPE), and, when applicable, Bayes factors.  While \textit{algorithmic} assessment is not required in \pkg{bayesics}, \textit{model} assessment is still critical; towards that, \pkg{bayesics} provides diagnostic plots for parametric inference, including Bayesian p-values. Finally, \pkg{bayesics} provides extensions to models implemented in alternative \R packages and, in the case of mediation analysis, correction to existing implementations.
}

\Plainkeywords{Bayesian inference, Bayesian non-parametrics, diagnostic plots, one sample inference, R, regression, two sample inference}

\Address{
  Daniel K. Sewell\\
  Department of Biostatistics\\
  University of Iowa\\
  145 n. Riverside Dr., Iowa City, IA 52242, USA\\
  E-mail: \email{daniel-sewell@uiowa.edu}\\
  URL: \url{https://sewell.lab.uiowa.edu/}
}

\begin{document}
\Sconcordance{concordance:bayesics-jss-sub1.tex:bayesics-jss-sub1.Rnw:1 44 1 1 10 72 1 1 40 %
154 1 1 33 1 5 6 1 1 57 1 2 42 1 1 2 1 0 1 1 1 11 13 0 1 2 4 1 1 5 4 0 1 4 6 0 1 %
2 6 1 1 5 4 0 1 4 3 0 1 2 4 0 1 2 13 1 1 3 2 0 1 4 2 0 1 3 2 0 1 2 4 0 1 2 4 1 1 %
2 5 0 1 2 12 1 1 2 1 0 1 3 2 0 1 4 2 0 1 4 6 0 1 2 5 1 1 13 16 0 1 2 59 1 1 3 2 %
0 1 2 1 4 37 0 1 2 14 1 1 4 3 0 1 4 6 0 1 2 10 1 1 5 8 0 1 2 9 1 1 5 4 0 1 3 5 0 %
1 2 5 1 1 2 25 0 1 2 7 1 1 3 6 0 1 2 21 1}



\section[Introduction]{Introduction}
\label{sec:intro}

As is done colloquially, Bayesian statistics equates \textit{probability} with \textit{uncertainty}.  Subsequently, a Bayesian definition of probability intends to quantify the amount of uncertainty, or lack of knowledge, about a particular truth or event.  Bayesian analyses attempt to directly answer scientific queries, such as, ``Is my hypothesis true?'' or ``How certain are we that there is a positive relationship between $x$ and $y$?'', leading to directly interpretable quantities. In contrast, frequentist inference is often difficult to interpret or requires additional steps to make scientific conclusions.  For example, confidence intervals only allow for interpretation about the process and not about any specific realized confidence interval. 
The other frequentist mainstay is p-values, which are ``too often misunderstood and misused in the broader research community'' \citep{asapvalues} and have contributed to significant reproducibility concerns \citep{colquhoun2017reproducibility}. Indeed, \cite{greenland2016statistical} lists a large number of ways frequentist estimates and procedures are commonly misinterpreted, many of which correspond to correct interpretations from a Bayesian framework.

\begin{figure}
\centering
\includegraphics[width=0.75\textwidth]{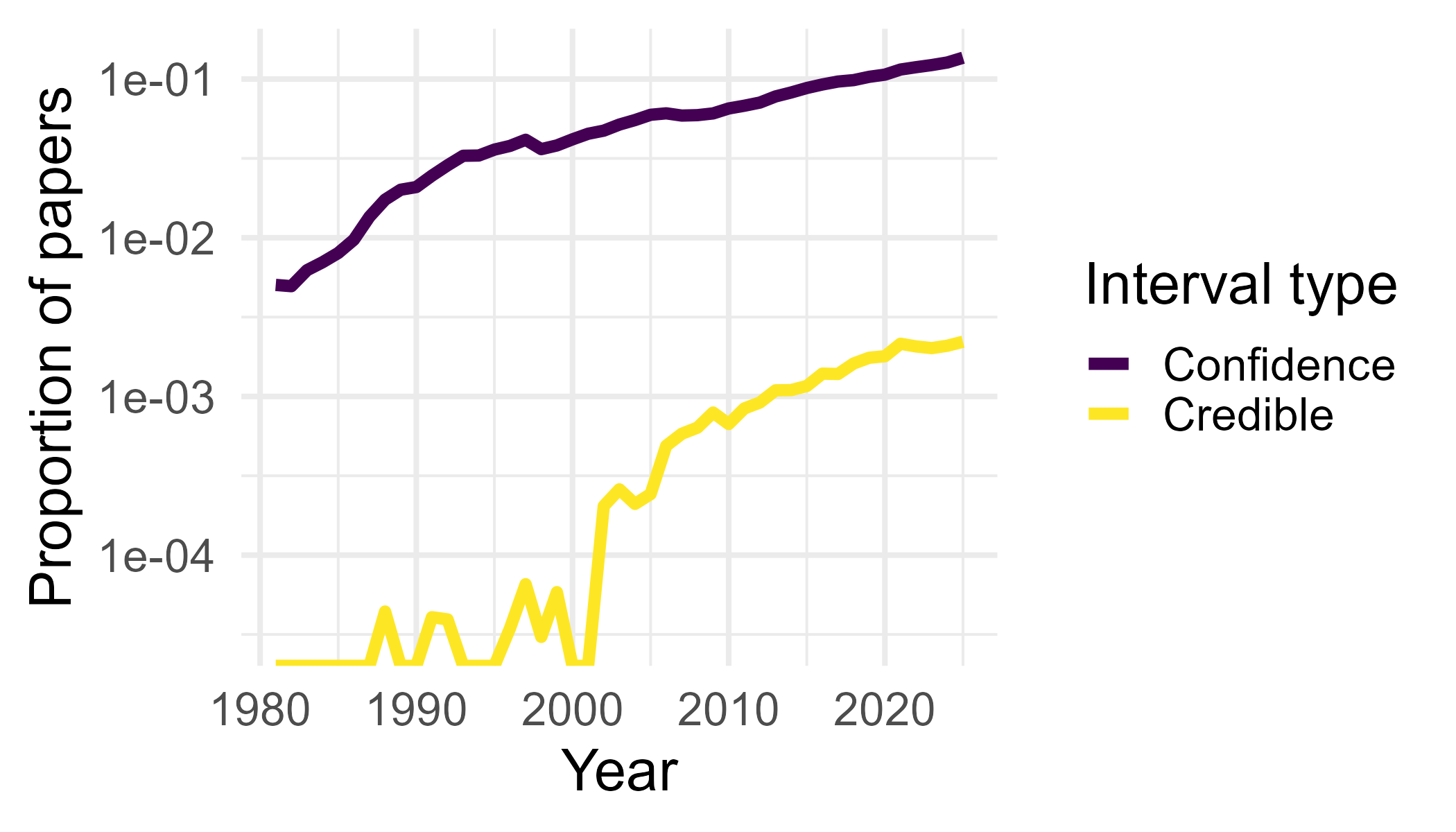}
\caption{Proportion of peer-reviewed articles catalogued by \textit{Europe PMC} that contain either ``credible interval'' or ``confidence interval'' from 1981 - 2025.}
\label{fig-europepmc}
\end{figure}

Thanks to prominent figures (de Finetti, DeGroot, Jeffreys, Savage, etc.) the Bayesian paradigm is well founded philosophically.  It has a strong grounding in both theory and computation, is embedded in contemporary applied science and machine learning \citep{bon2023being}, and is becoming more widely accepted in the life sciences by both industry and governmental regulatory bodies \citep{rosner2021bayesian}. Figure \ref{fig-europepmc} shows the proportion of all peer-reviewed articles catalogued by \textit{Europe PMC} that include the phrase ``credible interval'' as a proxy (and at least a lower bound) for how often Bayesian inference is being conducted in biomedical research.  Starting around 2003, Bayesian inference has become drastically more widespread.  However, Figure \ref{fig-europepmc} also shows the trend for the search term ``confidence interval,'' illustrating that despite the dramatic uptick in the use of Bayesian statistics, frequentist statistics still dominate the life sciences.  While the reasons for this are well beyond the scope of this paper, the authors conjecture that a primary driver may be the circularity of science and teaching described in the opening of \cite{asapvalues}- frequentist statistics is taught because that is what regulators and journal editors expect, and regulators and editors expect frequentist statistics because that is what they were taught.  On the other hand, the increased usage of Bayesian statistics has been asserted to be in large part the product of increased computational tools \citep[e.g., ][]{hagan2010oxford} and statistical software implementing Bayesian methodology \citep[e.g., ][]{strumbelj2024past}.  

There are a large number of \R packages available for obtaining Bayesian inference for very specific tasks, many of which correspond to advanced statistical methods (causal analysis, network analysis, spatial statistics, etc.); for example, \pkg{bama} \citep{bama} has the sole purpose of performing Bayesian inference for high-dimensional linear mediation models.  On the other extreme lie advanced \R packages aimed at building models from scratch and then obtaining Bayesian inference using them, most notably \pkg{greta} \citep{greta}, \pkg{LaplacesDemon} \citep{laplacesdemon}, \pkg{NIMBLE} \citep{devalpine2017programming}, and \pkg{rstan} \citep{rstan}.

More closely related to the aims of the newly developed \R package \pkg{bayesics}, the subject of this manuscript, are those \R packages which aim to do a set of commonly implemented analyses using Bayesian inferential techniques.  \pkg{BayesFactor} \citep{BayesFactor} can perform Bayesian hypothesis testing for ANOVA, linear regression, and correlation (for two normally distributed samples), but, as the name suggests, provides only Bayes factors, and doesn't provide point or interval estimation.  For common one- and two-sample inferential procedures, \pkg{DFBA} \citep{dfba} provides non-parametric Bayesian approaches, but no regression methods.  In contrast, \pkg{arm} \citep{arm} provides only regression methods.  While \pkg{arm} relies primarily on the large sample normal approximation for posterior inference, 
more recent \R packages that are focused on regression rely on \proglang{Stan}, a probabilistic programming language for specifying statistical models used primarily to implement No-U-Turn Hamiltonian Monte Carlo sampling \citep{carpenter2017stan}. Two of the most notable such \R packages are \pkg{rstanarm} which has a large array of precompiled statistical regression models \citep{rstanarm}, and \pkg{brms} which provides a highly flexible approach to building complex regression and hierarchical models \citep{brmsjss}.  These two packages are very powerful and useful for regression methods, yet it is all too easy to use these packages without obtaining reliable inference. As discussed in Section \ref{subsec-mcaccuracy}, knowing the Markov chains have converged is insufficient for determining if one has reliable inference, and the Monte Carlo standard error (MCSE) is quite misleading in this regard as it only pertains to the point, rather than interval estimate.  Furthermore, these packages do not automatically provide the critical inferential quantities needed to perform a Bayesian analysis (see Section \ref{subsec-inference}).  It should be noted that this has largely been rectified by the well-developed \R package \pkg{bayestestR} \citep{bayestestR}, although some issues remain with the default ROPE bounds, namely adjusting according to the scale of the covariates.

To address these issues, we have developed the \pkg{bayesics} package.  As its name suggests, its overall goal is to provide a single framework, unified in syntax and output, for performing the most commonly used statistical analyses using Bayesian inference.  The analyses provided in \pkg{bayesics} range from simple one- and two-sample analyses to somewhat more advanced regression techniques such as model averaging and general mediation analysis.  Table \ref{tbl-functionality} provides a full list of the functionality of \pkg{bayesics}.  Beyond this, one of the most important motivations for \pkg{bayesics} was to emphasize inference over algorithms, and to provide ways to better understand/interpret the output, including visualization tools.  It seems non-controversial that to perform good science, it is sufficient to be able to use and correctly interpret the output of, e.g., \code{stats::glm()} without in-depth knowledge of how \code{stats::glm()} obtains its results via iteratively reweighted least squares.  Thus, the authors feel strongly that Bayesian inferential procedures ought to be accessible in the same way.  Towards that, \pkg{bayesics} implements methods that either require no algorithmic tuning (i.e., closed-form solutions) or do so automatically with minimal input from the user (namely, accuracy tolerance).  Importantly, we focus on the interval estimates rather than the point estimates, as these are often, rightly or wrongly, what are used to determine a ``statistically significant'' result (e.g., if the 95\% credible interval for a regression coefficient does not include zero) and require considerably more posterior samples to obtain sufficient precision.  \pkg{bayesics} also provides diagnostic plots for model assessment (none are needed for algorithmic assessment) including the implementation of Bayesian p-values for generalized linear models (GLMs), as well as non-parametric alternatives to regression models should model assumptions fail.  Finally, \pkg{bayesics} provides extensions to models implemented in alternative \R packages and, in the case of mediation analysis, corrections to existing methods.

In the remainder of the paper, we go into more detail on the emphasis on inference and understanding over algorithms (Section \ref{sec-inference}), model diagnostic plots and remediations (Section \ref{sec-dx}), and finally method extensions and corrections (Section \ref{sec-ext}).  While Sections \ref{sec-inference}-\ref{sec-ext} include a few examples of various \pkg{bayesics} functions, Section \ref{sec-demo} provides a fuller demonstration. We end with a discussion in Section \ref{sec-discussion}.

\begin{longtable}{p{4cm}p{5cm}p{3.5cm}}
	\caption{Functions, analyses, and available generics. Dependencies are noted in the footnotes.} \label{tbl-functionality} \\
	\hline
	\textbf{Function} & \textbf{Analysis} & \textbf{Generics}  \\
	\hline
	\endfirsthead

	\hline
	\textbf{Function} & \textbf{Analysis} & \textbf{Generics}  \\
	\hline
	\endhead

	\code{aov\_b} & 1-way analysis of variance &
	  \code{coef}, \code{credint}$^1$, \code{*IC}$^2$, \code{plot}, \code{predict}, \code{print}, \code{summary}, \code{vcov} \\
	\hline
	\code{bma\_inference}$^3$ & Bayesian model averaging for linear regression models & \code{coef}, \code{credint}, \code{plot}, \code{predict}, \code{print}, \code{summary} \\
	\hline
	\code{case\_control\_b} & Analysis of 2x2 contingency tables from a case-control study & \\
	\hline
	\code{chisq\_test\_b} & Independence analysis for 2-way contingency tables & \\
	\hline
	\code{cor\_test\_b}$^4$ & Analysis of Kendall's tau correlation coefficient & \\
	\hline
	\code{find\_beta\_parms} & Find shape parameters of the Beta distribution which tries to match a user-specified mean and quantile & \\
	\hline
	\code{find\_invgamma\_parms} & Find shape and rate parameters of the Inverse Gamma distribution that either tries to match two user-specified quantiles, or else finds an Inverse Gamma distribution that matches a priori confidence in the coefficient of determination ($R^2$) & \\
	\hline
	\code{glm\_b} & Generalized linear models & \code{bayes\_factors}, \code{coef}, \code{credint}, \code{*IC}, \code{plot}, \code{predict}, \code{print}, \code{summary}, \code{vcov} \\
	\hline
	\code{heteroscedasticity\_test} & Test whether two or more populations have equal variances via Bayes factors & \\
	\hline
	\code{lm\_b} & Linear models &
	\code{bayes\_factors}, \code{coef}, \code{credint}, \code{get\_posterior\_draws}, \code{*IC}, \code{plot}, \code{predict}, \code{print}, \code{summary}, \code{vcov} \\
	\hline
	\code{mediate\_b} & Mediation analysis using the framework of \cite{imai2010general} &  \code{plot}, \code{print}, \code{summary}\\
	\hline
	\code{np\_glm\_b} & Generalized Bayesian inference via the loss-likelihood bootstrap &
	\code{coef}, \code{credint}, \code{plot}, \code{predict}, \code{print}, \code{summary}, \code{vcov} \\
	\hline
	\code{poisson\_test\_b} & Make inference on one or two populations using Poisson distributed count data & \\
	\hline
	\code{prop\_test\_b} & Make inference on a single population proportion or compare two population proportions & \\
	\hline
	\code{sign\_test\_b} & Sign test for paired data & \\
	\hline
	\code{surv\_fit\_b} & Semi-parametric estimation of survival curves for right-censored data for one or more populations & \code{bayes\_factors}, \code{plot}, \code{print}\\
	\hline
	\code{t\_test\_b} & Make inference on one or two population means using normally distributed data & \\
	\hline
	\code{wilcoxon\_test\_b}$^5$ & Wilcoxon Rank Sum (aka Mann-Whitney U) or the Wilcoxon signed rank analyses & \\
	\hline
	\multicolumn{3}{l}{$^1$ \code{credint()} is the Bayesian equivalent of \code{stats::confint()}} \\
	\multicolumn{3}{l}{$^2$ *IC represents the following information criteria: AIC, BIC, DIC, and WAIC} \\
	\multicolumn{3}{l}{$^3$ Depends on \code{BMS::bms}} \\
	\multicolumn{3}{l}{$^4$ Depends on \code{DFBA::dfba\_bivariate\_concordance}} \\
	\multicolumn{3}{l}{$^5$ Depends on \code{DFBA::dfba\_mann\_whitney}, \code{DFBA::dfba\_wilcoxon}} \\

\end{longtable}


\section[Inference, not Algorithms]{Inference, not Algorithms}
\label{sec-inference}

\subsection[What can Bayesians do for you?]{What can Bayesians do for you?}
\label{subsec-inference}

We assume that the reader has at least a reasonable familiarity with Bayesian statistics. For those who need a primer, there exist countless introductory texts on the subject \citep[e.g.,][]{rosner2021bayesian}.  For those well versed in Bayesian analysis, we recommend skipping to Section \ref{subsec-mcaccuracy}.  

Before proceeding, we will first cover at a very high level the fundamental inferential quantities used by the practicing Bayesian statistician, as these are referred to throughout the remainder of the manuscript. The first component is, of course, the point estimate.  Typically used is the posterior mode, posterior median, or the posterior mean.  Each of these has a strong theoretical justification rooted in decision theory, but it is sufficient here to say that our posterior mean, for example, is our best guess at the truth.  This has very limited practical implications, however, as our best guess is undoubtedly wrong\footnote{Unless our estimand takes one of a discrete set of values.}.  

Our second key inferential quantity is credible intervals (CIs), which provide a better approach to understanding what the truth might be.  CIs are precisely what practitioners wish confidence intervals were, and unfortunately, the latter are often misinterpreted as the former \citep[this is, in fact, the very first misinterpretation of confidence intervals listed by ][]{greenland2016statistical}.  If a Bayesian analysis provides a 95\% CI for an estimand, we say correctly that we are 95\% confident that the truth lies in our interval.

Third, when we are performing a causal or association study, we wish to quantify our certainty that we understand the direction of the relationship.  In regression analysis, this corresponds to statements of confidence about whether the regression coefficient is, e.g., positive.  This quantity is called, appropriately enough, the probability of direction, or simply PDir.  

While the PDir is important for characterizing our knowledge of the direction of a relationship, it is wholly insufficient for telling us whether that relationship is practically meaningful.  To address this, practitioners ought to define a region of practical equivalence (ROPE). With this, a Bayesian analysis can determine how certain we are that the estimand lies within this ROPE. \cite{kruschke2018rejecting} provides a well thought out discussion on default ROPE bounds, which \pkg{bayesics} adheres to for default settings.

Finally, we sometimes wish to compare the likelihood of one statement's fidelity against that of a contradictory statement.  For example, we may wish to compare head-to-head the statement, ``exposure to chemical A is associated with an increased chance of disease B,'' with the statement ``exposure to chemical A is \textit{not} associated with disease B.''  Bayes factors tell us how our prior beliefs about the veracity of these two competing statements change based on the data, regardless of what those prior beliefs are.  More specifically, the Bayes factor will scale the prior odds that statement 1 is true vs. statement 2 is true to obtain these same odds after having seen the data.  

These are the most well used and important pieces of a Bayesian analysis, and it is the goal of \pkg{bayesics} to provide and facilitate interpretation of these pieces while allowing the user to neglect the method of how these quantities were obtained algorithmically.

\subsection{Monte Carlo accuracy}
\label{subsec-mcaccuracy}
Bayesian inference is often performed in practice through posterior sampling- usually MCMC.  Thus it is important to understand the level of precision obtained by a specified number of posterior samples.  Most implemented algorithms use some measure of convergence diagnostic, e.g., the Geweke diagnostic \citep{geweke1991evaluating} or $\hat R$ \citep{gelman2004BDA3}, and the Monte Carlo standard error (MCSE) for the posterior mean.  The former is clearly critical to know whether the samples are valid, and the latter is intended to indicate whether or not sufficient numbers of valid samples have been obtained.  Yet the authors suspect that unless the user is well versed in posterior sampling and MCMC, the former measures of chain convergence are used to determine whether the reported inference is accurate.  However, even were the MCSE to be used to determine whether sufficient samples were obtained, this measure can be highly misleading for anything other than point estimation.

The important work in \cite{doss2014markov} provides a central limit theorem for order statistics for MCMC-derived samples.  For explication purposes, we will use the ideal setting of \textit{iid} samples, although the problem becomes further exacerbated by high autocorrelation in the chain.  For a parameter of interest $\theta$, let $\hat\theta_{L,p}$ denote the $p^{th}$ empirical quantile of an \textit{iid} sample of size $L$ (note that $L$ represents the number of posterior samples, not the sample size of the original data), let $\theta_{p}$ denote the corresponding true posterior quantile, and let $\pi(\cdot|y)$ denote the posterior probability density function of $\theta|y$.  Then
\begin{equation}
  \sqrt{L}(\hat\theta_{L,p} - \theta_{p}) \overset{d}{\to} N\left(0,\frac{p(1-p)}{\pi^2(\theta_p|y)}\right).
\end{equation}
Hence, to estimate, say, the lower bound of a $(1-\alpha)100$\% CI, i.e., the $(\alpha/2)^{th}$ posterior quantile of $\theta|y$, up to $\pm$ some error $\epsilon$ with (high) probability $s$, i.e.,
$$
s = \Pr(\theta_{\frac{\alpha}{2}} - \epsilon < \hat\theta_{L,\frac{\alpha}{2}} < \theta_{\frac{\alpha}{2}} + \epsilon),
$$
we set $L$ to be
\begin{equation}
  L =
  \frac{\alpha}{2} \left(1 - \frac{\alpha}{2}\right)
  \left(
    \frac{Z_{\frac{1-s}{2}}}{\epsilon \pi(\theta_{\frac{\alpha}{2}}|y)}
  \right)^2
  \label{eq-dosssamplesize}
\end{equation}

The MCSE comes into play when determining the number of samples required to obtain accurate estimation of the posterior mean.  Let $\bar\theta$ denote the true posterior mean and $\hat{\bar \theta}_M$ denote the estimate of the posterior mean from $M$ \textit{iid} posterior samples. Then to estimate $\bar\theta$  up to $\pm\epsilon$ with probability $s$, the number of samples required is
\begin{equation}
  M =
  \text{Var}\left(\theta^{(m)}\right)\left(
    \frac{Z_{\frac{1-s}{2}}}{\epsilon}
  \right)^2.
\end{equation}
where $\theta^{(m)}$ is the $m^{th}$ \textit{iid} posterior draw of $\theta$. Putting these together, we quickly see that for whatever probability $s$ and Monte Carlo (MC) margin of error $\epsilon$, the ratio of samples required to obtain such precision for the lower credible interval bound to that required for the posterior mean is given by
\begin{equation}
  \frac{\frac{\alpha}{2}\left(1 - \frac{\alpha}{2}\right)}{\text{Var}\left(\theta^{(m)}\right) \pi^2\left(\theta_{\frac{\alpha}{2}}|y\right) }
\end{equation}

\begin{figure}[h]
\centering
\includegraphics{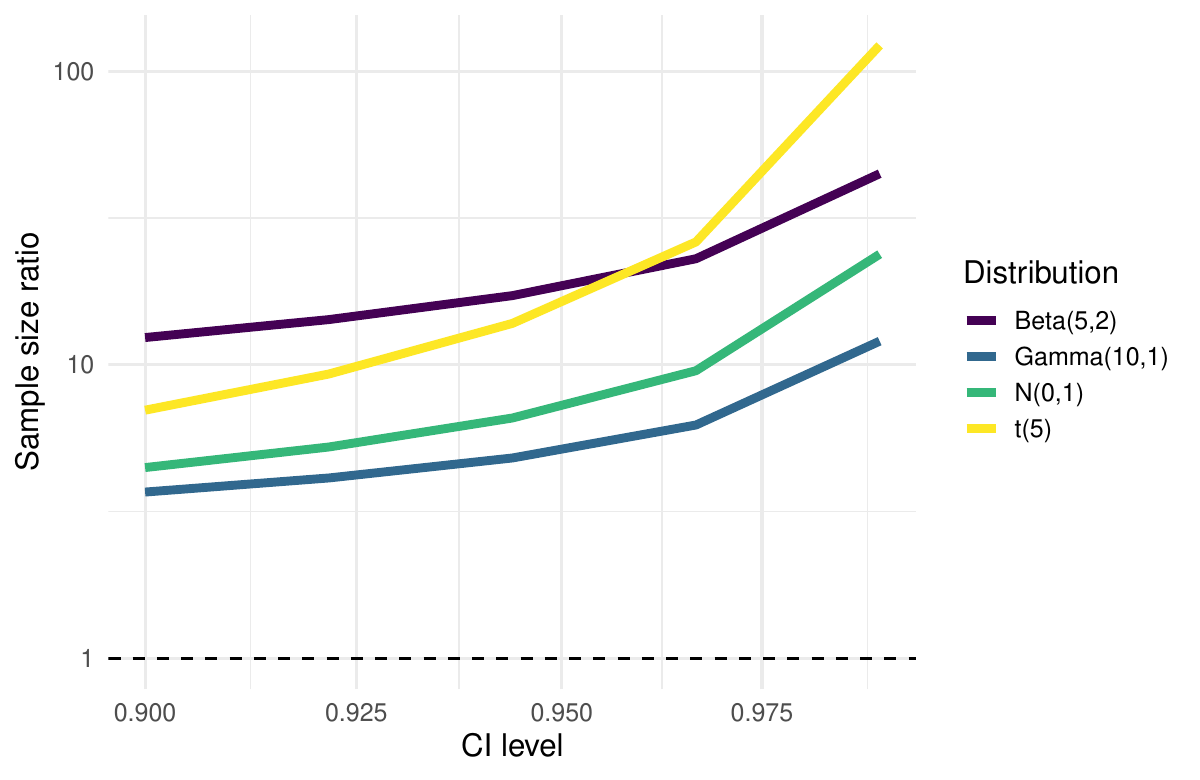}

\caption{Illustration of the ratio of posterior samples required for the lower credible interval bound vs. the posterior mean to be within the same margin of error.  CI levels range from 90\% to 99\%.}
\label{fig-samplesizeratio}
\end{figure}

Figure \ref{fig-samplesizeratio} illustrates the egregious nature of this problem when considering how many more samples are required for tail quantile estimation than for posterior mean estimation.  From this, we can see there is the potential for several orders of magnitude more samples to be required for equivalent precision in the interval estimates as for the point estimates.  To further illustrate this, we simulated data according to a simple linear regression model with intercept $=0$, slope $=0.25$, and residual standard deviation$=1$, and a sample size of 25.  We then employed \pkg{rstanarm}, using their default settings, to estimate using 500 simulations the standard error for the lower bound of the 95\% credible interval for the slope and compared that with the MCSE reported for the posterior mean.  Across 500 simulations, the reported MCSE was consistent, ranging from 0.0034 to 0.0047.  However, the empirical standard error of the lower bound of the 95\% credible interval was 0.0133, roughly 3-4 times the error reported by the MCSEs.

\FloatBarrier
\FloatBarrier

\subsection[bayesics automated approach]{\pkg{bayesics}  automated approach}

When possible, \pkg{bayesics} avoids the requirement for Monte Carlo numerical approximation by using conjugacy.  A simple example of this is \code{lm\_b}, which uses a normal-inverse gamma prior on the regression coefficients and the residual variance for linear regression.  That is, for the familiar model given by $\bv{y} = X\bv\beta + \bv\epsilon$, we use the priors

\begin{equation}
    \bv\beta|\sigma^2 \sim N(\bv\mu,\sigma^2 V^{-1}), \qquad
    \sigma^2 \sim \Gamma^{-1}(a/2,b/2),
\end{equation}
which yields a multivariate t distribution for the marginal prior and marginal posterior for $\bv\beta$. The default setting in \pkg{bayesics} is a Zellner's $g$ prior with $g$ equalling the sample size. Other priors are available, notably that which corresponds to the statement, ``we are 95\% sure a priori that a standard deviation increase in any covariate will not change the mean of $y$ by more than a factor of 5 standard deviations of $y$,'' (obtained when \code{prior="conjugate"}).  The default hyperparameters for $\sigma^2$ are motivated by placing a prior on the coefficient of determination $R^2$.  Specifically, since the residual variance is $(1-R^2)$ times the observed variance, we set 50\% prior probability that $\sigma^2$ is between $(1-0.9^2)s_y^2$ and $(1-0.1^2)s_y^2$, where $s_y^2$ is the sample variance of the response variable, i.e., we are 50\% confident a priori that the correlation between the observed  and the fitted values will be between 0.1 and 0.9 in magnitude.

In the case of GLMs, no conjugate prior exists.  Instead, \pkg{bayesics} implements through \code{glm\_b} a fixed form variational Bayes (VB) (as well as importance sampling and large sample approximations, although VB is the default), which finds the multivariate normal distribution with unconstrained posterior covariance matrix which is as close as possible to the posterior with respect to the Kullback-Leibler divergence \citep{salimans2013fixedform}.  We note that this approach is also implemented in, e.g., \code{rstanarm::stan\_glm} if one sets the argument \code{algorithm="fullrank"}.  

In other cases where there is no conjugate prior available, obtaining \textit{iid} posterior samples is trivially accomplished, and in such cases \pkg{bayesics} automatically determines the number of posterior draws.  For example, suppose we wish to study the rate ratio from two independent samples which are counts, i.e., we have $y_j\sim Poisson(\lambda_j\omega_j)$, $j=1,2$, where $\lambda_j$ is the population rate and $\omega_j$ is the offset term. Using a conjugate gamma prior for each individual population rate allows large numbers of posterior draws of $\lambda_j|y_j$, and thus posterior draws of $\lambda_1/\lambda_2 | y_1,y_2$, to be obtained trivially.  The \code{poisson\_test\_b} draws a preliminary sample of size 500 which is then used to estimate the density at the desired quantile corresponding to the lower and upper bounds of the credible interval. This is the key quantity in determining the number of posterior samples as seen in Equation \ref{eq-dosssamplesize}, and \code{poisson\_test\_b} will automatically draw the remaining samples necessary for the desired level of precision. 

Sometimes a mix of these two approaches- closed-form solutions and sampling- is implemented.  In \code{aov\_b}, which implements one-way ANOVA, some estimands such as individual factor level means have closed-form posterior distributions, while others such as exceedance in pairs rate (EPR)\footnote{EPR is a quantity described in \cite{rosner2021bayesian} p.136 measuring the probability that a random value from one population is greater than a random value from another population, i.e., how well separated the two population distributions are.} do not.  For these latter cases we can still easily obtain \textit{iid} posterior samples and, as with \code{poisson\_test\_b}, the number of such draws is determined in an automated fashion.  In this way, \textbf{the user of \pkg{bayesics} should very rarely need to be concerned with any aspect of the algorithms used, and can trust that the output will be correct and accurate.}

\section[Diagnostics and Remediations]{Diagnostics and Remediations}
\label{sec-dx}

\subsection{Diagnostic plots}
While diagnostic plots relating to algorithms, e.g., MCMC trace plots, are not required by \pkg{bayesics} algorithms, it is still crucial to assess a model's goodness of fit.  Towards that, for one-way ANOVA (\code{aov\_b}) and linear regression (\code{lm\_b}) models, the \code{plot} function provides the standard residual vs. fitted values plot as well as a qq-norm plot (among other options).  

For GLMs, it is just as important to assess model fit.  \pkg{bayesics} addresses this need through the use of Bayesian p-values. Briefly, Bayesian p-values, also called posterior predictive p-values, assess the compatibility of a certain feature of a model to the observed data.  Letting $y_{obs}\in{\cal Y}$ denote the observed data, $y_{pred}\in{\cal Y}$ denote data coming from the posterior predictive distribution, $\theta\in\Theta$ be the proposed model's parameters, and $T:{\cal Y}\times \Theta \mapsto \Re$ be the test statistic, the Bayesian p-value is defined to be

\begin{equation}
  \Pr(T(y_{pred},\theta) < T(y_{obs},\theta) | y_{obs}).
\end{equation}
Note that the test statistic can be a function of both the observables and the model parameters.  If the model is correct, or at least correct in terms of what the test statistic $T(\cdot,\cdot)$ is assessing, then the model ought to predict values similar to the observed data.  In such a case, we would hope that the Bayesian p-value would near 0.5.  If, however, we obtain a p-value near 0 or 1, say, $<0.05$ or $>0.95$, then we must conclude that the aspect of the model captured by our test statistic is inconsistent with the observed data.  For more details, see, e.g., \cite{gelman2004BDA3}.  \pkg{bayesics} provides scatterplots of posterior draws of $(T(y_{pred},\theta), T(y_{obs},\theta))$ pairs, and provides the p-value estimates. 

As a simple example, consider regression data generated according to the negative binomial distribution.  

\FloatBarrier
\begin{Schunk}
\begin{Sinput}
R> set.seed(2026)
R> N = 500
R> nb_data <- 
+    tibble(x1 = rnorm(N),
+           x2 = rnorm(N),
+           x3 = rep(letters[1:5],each = N/5),
+           time = rexp(N)) |> 
+    mutate(outcome =
+             rnbinom(N,
+                     mu = 
+                       exp(-2 + x1 + 2 * (x3 %in% c("d","e"))) *
+                       time,
+                     size = 0.7))
\end{Sinput}
\end{Schunk}
\FloatBarrier

\begin{figure}[h!]
\centering

\begin{Schunk}
\begin{Sinput}
R> p1 <- 
+    plot(poisson_fit,
+         type = "dx") +
+    labs(subtitle = "Poisson assumed")
R> p2 <-
+    plot(nb_fit,
+         type = "dx") +
+    labs(subtitle = "Neg. binomial assumed")
R> patchwork::wrap_plots(p1,p2)
\end{Sinput}
\end{Schunk}
\includegraphics{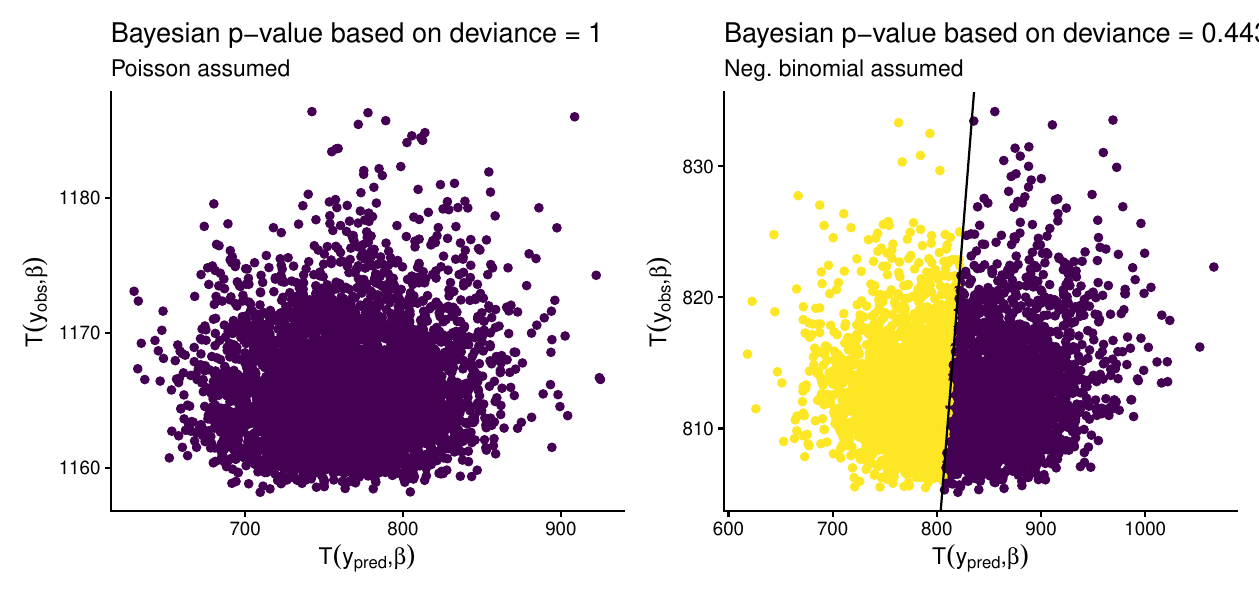}
\caption{Bayesian p-values for regression data generated according to the negative binomial}
\label{fig-pvals}
\end{figure}

If we fit a GLM based on the Poisson distribution, the overdispersion ought to be reflected in a p-value near 0 or 1, whereas when we fit a GLM based on the correct distribution\footnote{The \code{negbinom()} family is provided in \pkg{bayesics}.}, the p-value ought to be near 0.5.  Indeed, this is precisely what we see in Figure \ref{fig-pvals}.

\FloatBarrier
\begin{Schunk}
\begin{Sinput}
R> poisson_fit <-
+    glm_b(outcome ~ .,
+          data = nb_data,
+          family = poisson())
R> nb_fit <-
+    glm_b(outcome ~ .,
+          data = nb_data,
+          family = negbinom())
\end{Sinput}
\end{Schunk}
\FloatBarrier

\FloatBarrier
\FloatBarrier

Finally, Bayesian p-values are also implemented for Bayesian model averaging (more in Section \ref{subsec-bma}), where the test statistics used are quantiles of the data.

\subsection{Non-parametric methods}
If no candidate models are consistent with the observed data in terms of the Bayesian p-values or other diagnostic checks, it is better to opt for a non-parametric approach.  For simple scenarios such as paired analyses, we can use procedures such as the Bayesian sign test (\code{sign\_test\_b}).  The \pkg{DFBA} package provides good implementations of many such one and two sample techniques, and rather than reinvent the wheel, we have incorporated several of their functions into the common interface and the inferential output seen elsewhere in \pkg{bayesics}. 

\begin{figure}[h]
\centering
\begin{Schunk}
\begin{Sinput}
R> plot(GBSG2_fit_horTh)
\end{Sinput}
\end{Schunk}
\includegraphics{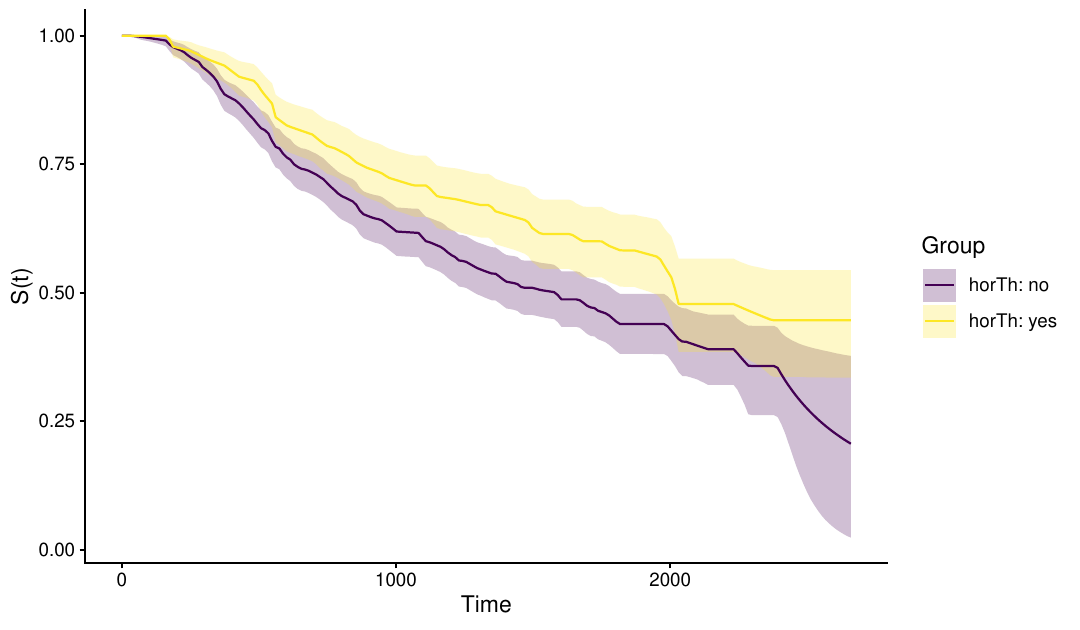}
\caption{Semi-parametric estimated survival curves for advanced lung cancer, disaggregated by sex.}
\label{fig-horThsurvfit}
\end{figure}

For fitting survival curves to one or more samples and/or testing group equivalence, we have implemented the semi-parametric approach of \cite{qing2023bayesian}.  In this setup, the hazard function is assumed to be a piecewise step function, or equivalently, the time-to-event data follow a piecewise exponential distribution.  This clever modeling approach is highly flexible, yet provides closed-form posteriors for the hazard rates, as well as a closed-form solution for the marginal likelihood.  The latter is important as it provides a ready approach for (1) determining the number of knots in the hazard function, and (2) testing whether all samples' time-to-event outcomes follow the same or different distributions via Bayes factors.  The code given below shows the analysis of survival times from breast cancer patients in the GBSG2 study (see \code{?TH.data::GBSG2} for more information and references), testing to see if survival rates vary by whether the patient received hormonal therapy, with the estimated survival curves given in Figure \ref{fig-horThsurvfit}.  Note that whenever \pkg{bayesics} provides Bayes factors, interpretations and direction are always made clear.

\FloatBarrier
\begin{Schunk}
\begin{Sinput}
R> data(GBSG2,
+       package = "TH.data")
R> GBSG2_fit_horTh <-
+    survfit_b(Surv(time,cens) ~ horTh,
+              data = GBSG2)
R> GBSG2_fit_aggregated <-
+    survfit_b(Surv(time,cens) ~ 1,
+              data = GBSG2)
R> bayes_factors(GBSG2_fit_aggregated,
+                GBSG2_fit_horTh)
\end{Sinput}
\end{Schunk}
\FloatBarrier

\FloatBarrier

For linear models and GLMs, \pkg{bayesics} implements the loss-likelihood bootstrap, a non-parametric method using general Bayesian updating \citep{lyddon2019general}.  Here, we consider the model parameters to be a functional of the true distribution of our observables involving a loss function, i.e., 
\begin{equation}
  \beta(F) := \underset{\tilde\beta}{\text{argmin}}\int_{{\cal Y}}\ell(\tilde\beta,y)dF(y),
\end{equation}
where $\beta(\cdot)$ is our functional parameter, $F$ is the true distribution of our observables $y\in {\cal Y}$, and $\ell(\cdot,\cdot)$ is our loss function mapping the parameter space and ${\cal Y}$ to the positive reals. 

Uncertainty in $\beta$ is solely the result of uncertainty in the distribution of $y$.  Thus, rather than needing to specify a prior on $\beta$, the loss-likelihood bootstrap uses the degenerate Dirichlet process prior on the distribution of the observables that corresponds to the Bayesian bootstrap.  Therefore the approach of \cite{lyddon2019general} has the arguable double benefit of not needing to specify a prior on the model parameters and not having to assume a correct likelihood.  As a default, the self-information loss is used (i.e., the loss function is the negative of the log likelihood), although any user-specified loss functions can be implemented.  To illustrate, consider a regression data with mis-specified error distribution and covariate structure.

\begin{figure}
\centering
\begin{Schunk}
\begin{Sinput}
R> plot(lm_fit,
+       type = "dx") +
+    (
+      plot(lm_fit,
+           type = "cred band") +
+        labs(subtitle = "(Parametric)")
+    ) +
+    (
+      plot(np_fit,
+           type = "cred band") +
+        labs(subtitle = "(Non-parametric)")
+    )
\end{Sinput}
\end{Schunk}
\includegraphics{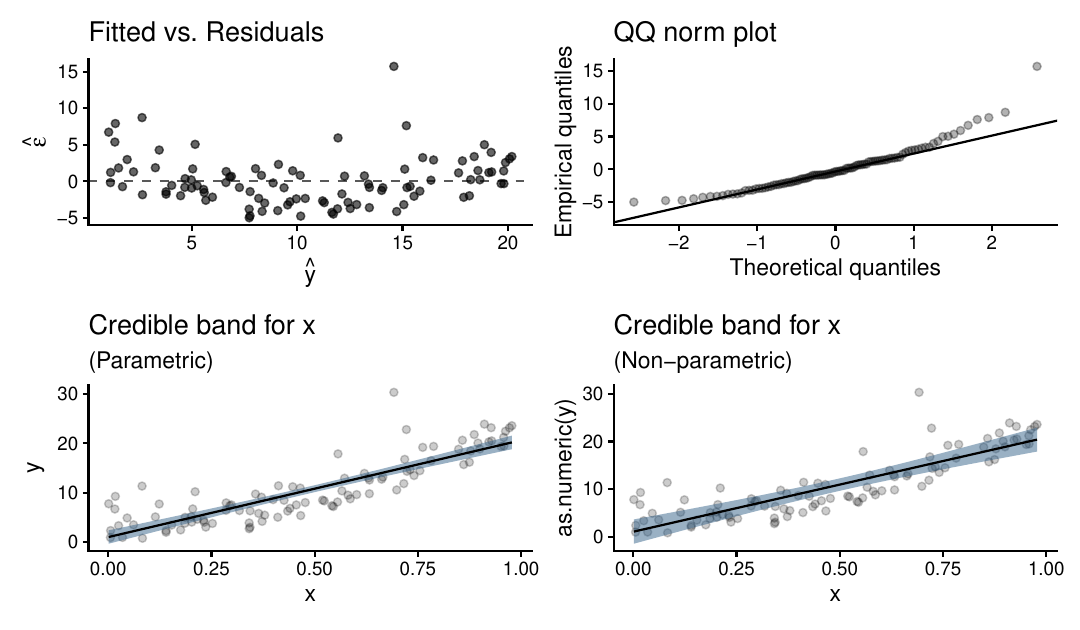}
\caption{A mis-specified linear regression model fit using a parametric approach and the non-parametric loss-likelihood bootstrap.}
\label{fig-npglm}
\end{figure}

\FloatBarrier
\begin{Schunk}
\begin{Sinput}
R> set.seed(2026)
R> linreg_data <-
+    tibble(x = runif(100)) |> 
+    mutate(y = 20 * x^2 + rgamma(100,shape = 2, rate = 0.5))
R> lm_fit <-
+    lm_b(y ~ x,
+         data = linreg_data)
R> np_fit = 
+    np_glm_b(y ~ x,
+             data = linreg_data,
+             family = gaussian())
\end{Sinput}
\end{Schunk}
\FloatBarrier
From Figure \ref{fig-npglm}, we can see that the diagnostic plots demonstrate violated assumptions (as expected), and the credible intervals are appropriately wider for the non-parametric compared to the parametric approach.  Note that when it comes to interpretation, the regression line estimated by \code{np\_glm\_b} is not a ``true'' regression line, but rather the regression line that would be fit were we to perfectly know the population distribution.

\FloatBarrier
\FloatBarrier

\section[Method Extensions and Corrections]{Method Extensions and Corrections}
\label{sec-ext}

\subsection[Information criteria]{Information criteria}
Information criteria play an important role in model selection and evaluation.  \pkg{bayesics} provides four of the most commonly used information criteria (IC): Akaike IC (AIC), Bayesian IC (BIC), Deviance IC (DIC), and Widely Applicable IC (WAIC).  Despite the popularity of these four criteria, with the exception of WAIC they are not all readily available as part of other widely used Bayesian packages.  In \pkg{bayesics}, all four are implemented for \code{aov\_b}, \code{lm\_b}, and \code{glm\_b}.

\subsection[Bayesian model averaging]{Bayesian model averaging}
\label{subsec-bma}
A powerful alternative to selecting and performing inference for a single model is Bayesian model averaging. This approach yields a mixture distribution as the posterior, where the mixture components are the posteriors obtained from each specific regression model and the component weights are the posterior probabilities that the corresponding models are correct.

The \pkg{BMS} package provides functionality to perform Bayesian model averaging.  Its eponymous function \code{bms} has considerable flexibility in terms of the prior on the model set, and the package provides many ways to describe the posterior over the model space and parameter inclusion probabilities.  Yet the capacity for inference on the coefficients themselves is minimal, which is where \pkg{bayesics} comes in.  \pkg{bayesics} obtains the posterior model probabilities from wrapping \code{BMS::bms}, and then leverages \code{lm\_b} to obtain inference.  Posterior sampling is required for this, but again \pkg{bayesics} automatically determines the necessary number of posterior samples to obtain accurate interval estimates.

\subsection[Mediation analysis]{Mediation analysis}
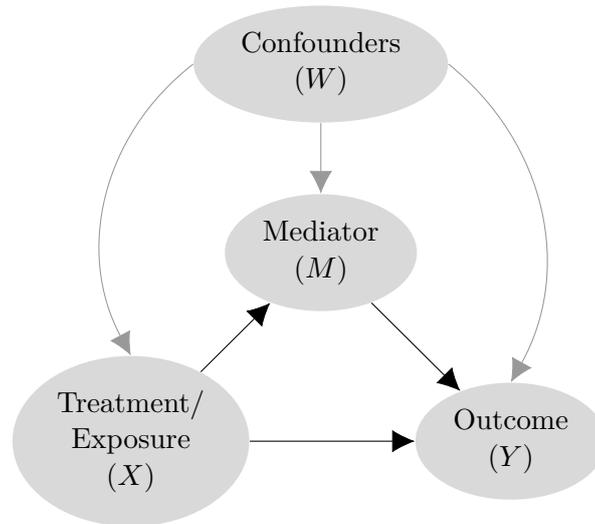
\begin{figure}
  \begin{center}
    \begin{tikzpicture}[scale = 5]
            \tikzstyle{every node} = [ellipse, fill = gray!30, align=center]
            \tikzstyle{every arrow} = [arrowhead = 10mm]
    
            \node (X) at (0,0) {Treatment/\\ Exposure\\ ($X$)};
            \node (Y) at (1,0) {Outcome\\($Y$)};
            \node (W) at (0.5,1) {Confounders\\ ($W$)};
            \node (M) at (0.5,0.5) {Mediator\\($M$)};
    
            \draw[->] (X) -- (Y);
            \draw[->] (X) -- (M);
            \draw[->] (M) -- (Y);
            \draw[->,gray!80] (W.west) to [bend right = 40] (X.north);
            \draw[->,gray!80] (W.east) to [bend left = 40] (Y.north);
            \draw[->,gray!80] (W.south) to (M.north);
            
        \end{tikzpicture}
  \end{center}
  \caption{Schematic of relationships assessed via a mediation analysis.}
  \label{fig-medschematic}
\end{figure}

Mediation analysis is an important method to gain a more nuanced understanding between predictor variables and an outcome of interest.  The type of relationships assessed in a mediation analysis is depicted in Figure \ref{fig-medschematic}.  \cite{imai2010general} described a highly flexible causal mediation framework in which the relationship between the treatment and the mediator and the relationship between the outcome and both the treatment and mediator can correspond to any sort of model.  However, their proposed estimation approach, so-called quasi-Bayesian, is lacking a rigorous framework.  In fact, their estimation approach corresponds to a Bayesian analysis using improper uniform priors.  The accompanying \R package \pkg{mediation}, however, refers to ``quasi-Bayesian confidence intervals'' and p-values.  In \pkg{bayesics}, we have provided fully Bayesian implementation of the causal mediation framework of \cite{imai2010general} for linear and generalized linear models.

\section[Demonstration]{Demonstration}
\label{sec-demo}
As a brief demonstration of \pkg{bayesics}, we consider the randomized clinical trial described in \cite{elmunzer2012randomized} studying the effects of indomethacin on post-endoscopic retrograde cholangio-pancreatogram pancreatitus (PEP).  These data are publicly available through the \R package \pkg{medicaldata}.  While this study contains a large number of variables, we focus only on a subset for pedagogical purposes.  Specifically, we consider as confounder variables age, sex, and previous history of PEP (found in the variable \code{risk}). The treatment arm is contained in the variable \code{rx} and equals 0 if placebo and 1 if treatment.  The binary outcome variable is contained in the variable \code{outcome} and indicates whether or not PEP occurred.  

While it not recommended in randomized trials to perform hypothesis tests to see if covariates are balanced, we shall do so here purely for demonstration purposes.  Age, treated as a continuous covariate, can be investigated via \code{t\_test\_b} after loading in the data.

\FloatBarrier
\begin{figure}[h]
\footnotesize 
\begin{Schunk}
\begin{Sinput}
R> data(indo_rct,
+       package = "medicaldata")
R> options(pillar.width = 100)
R> # Effective randomization for age
R> t_test_b(age ~ rx,
+           data = indo_rct)
\end{Sinput}
\begin{Soutput}
---
Bayes factor in favor of the full vs. null model: 1.21e-06;
      =>Level of evidence: Decisive



--- Summary of factor level means ---
# A tibble: 4 × 5
  Variable                   `Post Mean` Lower Upper `Prob Dir`
  <chr>                            <dbl> <dbl> <dbl>      <dbl>
1 Mean : rx : 0_placebo             46.0  44.6  47.5          1
2 Mean : rx : 1_indomethacin        44.5  42.9  46.0          1
3 Var : rx : 0_placebo             172.  147.  201.          NA
4 Var : rx : 1_indomethacin        183.  155.  215.          NA




--- Summary of pairwise differences ---
# A tibble: 1 × 9
  Comparison               `Post Mean`  Lower Upper `Prob Dir` `ROPE (0.1)`   EPR
  <chr>                          <dbl>  <dbl> <dbl>      <dbl>        <dbl> <dbl>
1 0_placebo-1_indomethacin        1.56 -0.461  3.64      0.933        0.412 0.533
  `EPR Lower` `EPR Upper`
        <dbl>       <dbl>
1       0.490       0.577


   *Note: EPR (Exceedence in Pairs Rate) for a Comparison of g-h = Pr(Y_(gi) > Y_(hi)|parameters) 
\end{Soutput}
\end{Schunk}
\includegraphics{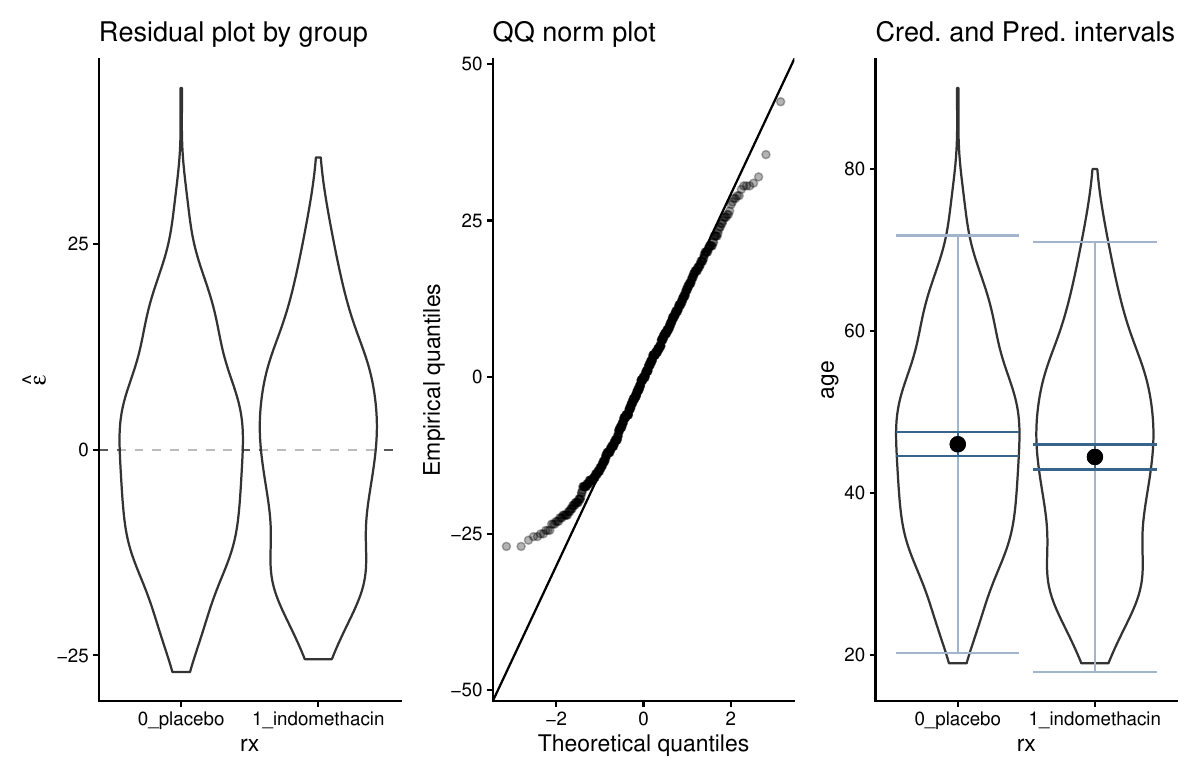}
\normalsize
\caption{Analysis of randomization of age between treatment arms.}
\label{fig-pepttest}
\end{figure}
\FloatBarrier

These results indicate effective randomization through the Bayes factor, credible interval of the difference in means, PDir, ROPE, and EPR.  It is important to recognize that
\code{t\_test\_b} automatically provides diagnostic plots, and in this case, the normality assumption is a little off due to the outcomes being right skewed.  If we were sufficiently concerned, we could instead run \code{wilcoxon\_test\_b} to implement the Bayesian rank-sum analysis (powered by \code{DFBA::dfba\_mann\_whitney}).

\begin{figure}[h]
\centering
\footnotesize
\begin{Schunk}
\begin{Sinput}
R> gender_table =
+    table(indo_rct$gender,
+          indo_rct$rx)
R> prop_test_b(gender_table[1,],
+              gender_table[2,],
+              CI_level = 0.99)
\end{Sinput}
\end{Schunk}
\includegraphics{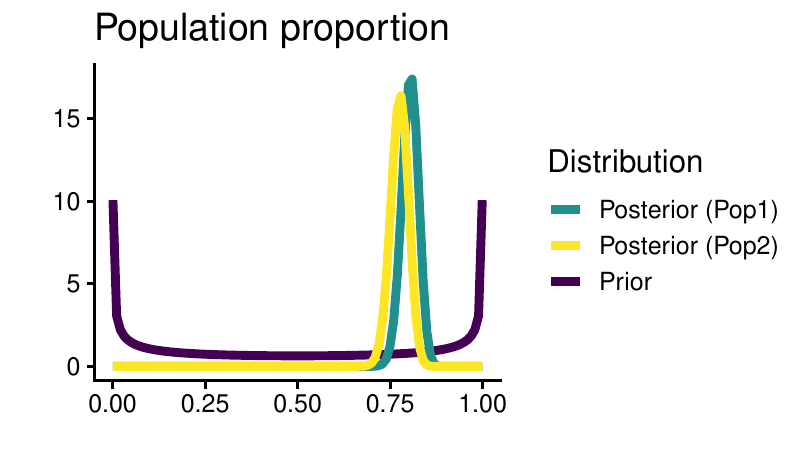}
\caption{Assessing randomization of sex (in this dataset, sex is binary) between treatment arms.}
\label{fig-pepptest}
\end{figure}

Next, we can assess whether or not sex was effectively randomized.  To achieve this, we can perform a simple test comparing two sample proportions.  \code{prop\_test\_b} provides the posterior means and credible intervals (for illustrative purposes, this time we elected to use 99\% CIs) for each population.  Additionally, the probability that the odds ratio is in the ROPE is provided, where the ROPE is, as suggested in the excellent discussion of \cite{kruschke2018rejecting}, half of a small effect size (per FDA guidance on rate ratios).  As seen in Figure \ref{fig-pepptest}, the default prior is a Jeffrey's prior, although this can be changed to a uniform prior by setting \code{prior="uniform"} or to a general Beta prior by using the \code{prior\_shapes} argument. 
\FloatBarrier
\FloatBarrier

Finally, we can check for randomization in the previous PEP \code{risk} variable, and this time, because \code{risk} is a discrete ordered risk score, we will use the Wilcoxon rank-sum analysis rather than \code{t\_test\_b}.  While the Bayes factor provides some evidence that randomization failed, the probability of direction and the probability of falling in the ROPE suggests this is not a major concern.

\FloatBarrier
\begin{figure}[h]
\centering
\footnotesize
\begin{Schunk}
\begin{Sinput}
R> wilcoxon_test_b(
+    indo_rct$risk[which(indo_rct$rx == "1_indomethacin")],
+    indo_rct$risk[which(indo_rct$rx == "0_placebo")]
+  )
\end{Sinput}
\end{Schunk}
\includegraphics{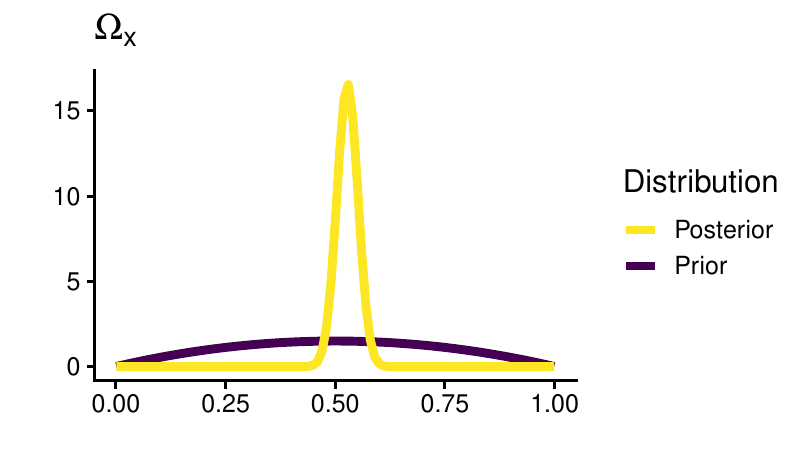}
\caption{Assessing randomization of risk from previous PEP between treatment arms.}
\label{fig-pepranksum}
\end{figure}
\FloatBarrier

We now move to the study's primary analysis (or at least, this pedagogical example's primary analysis): a covariate-adjusted generalized linear regression model to assess the effect of the indomethacin treatment on the PEP outcome. We fit the model using \code{glm\_b}, and then first look at the diagnostic plots.  

\FloatBarrier
\begin{figure}
\centering
\begin{Schunk}
\begin{Sinput}
R> pep_fit =
+    glm_b(outcome ~ age + gender + risk + rx,
+          data = indo_rct,
+          family = binomial())
R> plot(pep_fit,
+       type = "diagnostics")
\end{Sinput}
\end{Schunk}
\includegraphics{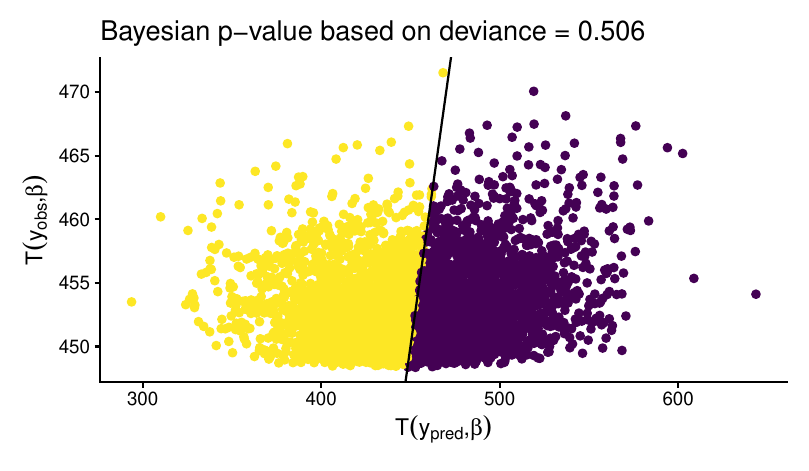}
\caption{Diagnostic plots for the primary PEP analysis.}
\end{figure}
\FloatBarrier

As we have a Bayesian p-value close 0.5, we can move on to inference. If the p-value were close to 0 or 1, we would have instead used \code{np\_glm\_b} with identical syntax.
\FloatBarrier
\begin{Schunk}
\begin{Sinput}
R> summary(pep_fit)
\end{Sinput}
\begin{Soutput}
----------

Values given in terms of odds ratios

----------

# A tibble: 4 × 9
  Variable         `Post Mean` Lower Upper `Prob Dir`    ROPE `ROPE bounds`
  <chr>                  <dbl> <dbl> <dbl>      <dbl>   <dbl> <chr>        
1 age                    0.993 0.975 1.01       0.754 0.145   (0.998,1.002)
2 gender2_male           1.10  0.611 2.00       0.629 0.288   (0.889,1.125)
3 risk                   1.53  1.17  2.00       0.999 0.00172 (0.967,1.034)
4 rx1_indomethacin       0.472 0.288 0.773      0.999 0.00572 (0.889,1.125)
  `BF favoring alternative` Interpretation                                   
                      <dbl> <chr>                                            
1                     0.157 Substantial  (in favor of exluding from the model
2                     0.129 Substantial  (in favor of exluding from the model
3                    14.3   Strong  (in favor of keeping in the model)       
4                    10.6   Strong  (in favor of keeping in the model)       
\end{Soutput}
\end{Schunk}
\FloatBarrier
The summary, unless the argument \code{interpretable\_scale} is set to \code{FALSE}, provides estimates in terms of the odds ratios.  From this, we see that there is an estimated 52.8\% (95\% CI: 22.7\% - 71.2\%) reduction in odds of PEP due to the treatment (\code{rx1\_indomethacin}).  The posterior probability that this effect is beneficial (i.e., a reduction in PEP) is very high (the probability of direction equals 0.999), and this is corroborated by strong evidence from the Bayes factor.  Moreover, this \textit{statistically} significant effect is also a \textit{practically} significant result, demonstrated by the posterior probability that the treatment effect falls in the ROPE is 0.006.  

\begin{figure}
\centering
\begin{Schunk}
\begin{Sinput}
R> plot(pep_fit,
+       type = "cred band")
\end{Sinput}
\end{Schunk}
\includegraphics{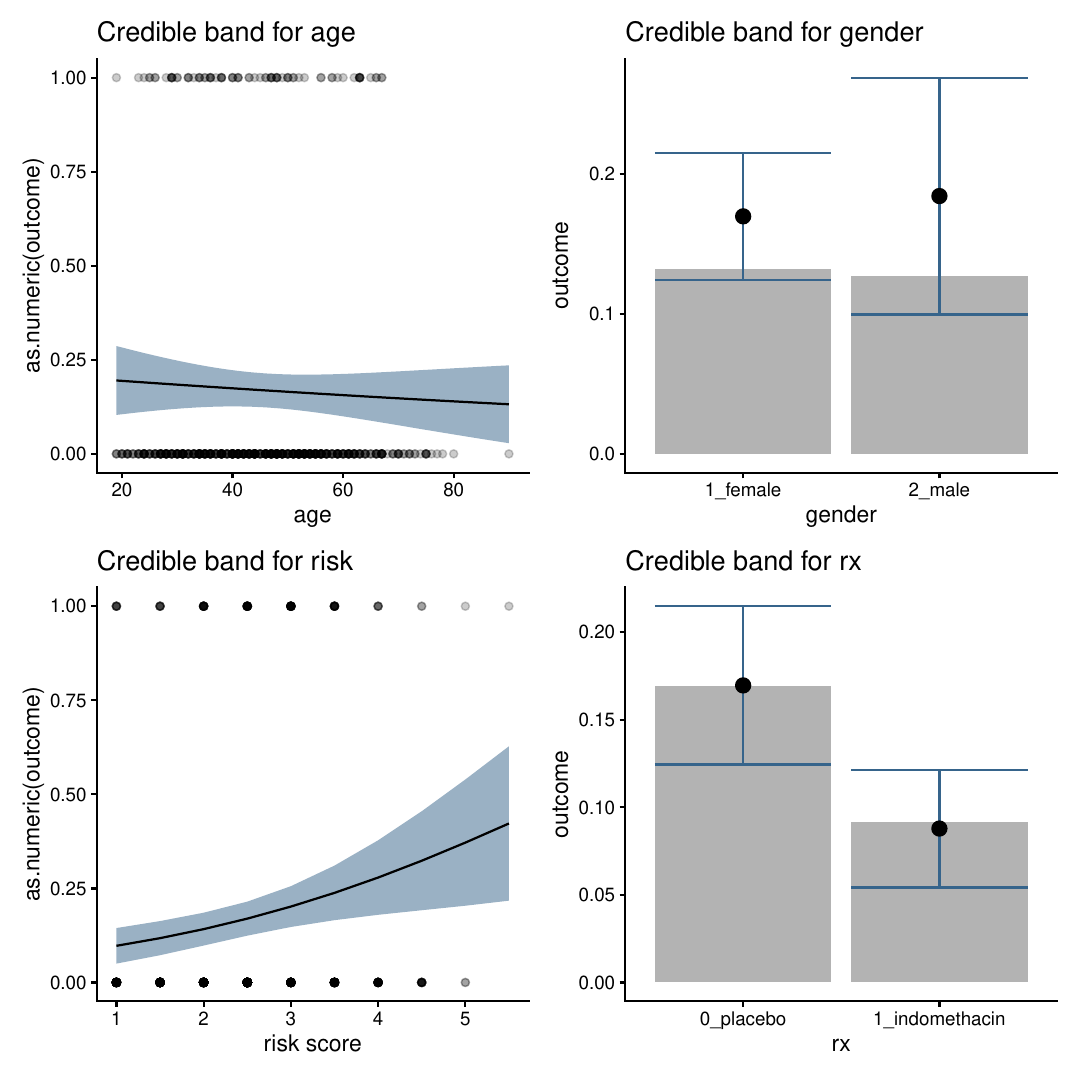}
\FloatBarrier
\caption{Credible bands for analyzing the PEP outcome using a GLM based on the Bernoulli distribution.}
\label{fig-pepbands}
\end{figure}

These effects can easily be visualized.  Figure \ref{fig-pepbands} shows the credible bands for the four covariates; note that these visualizations are given on the 0 to 1 scale for easier interpretation.  When \pkg{bayesics} creates such credible or prediction bands, other covariates must be fixed.  Users can fix these other values themselves via the \code{exemplar\_covariates}, but otherwise \pkg{bayesics} finds and uses the medoid observation from the data used to fit the model, ensuring that the covariate combination is a good representation of the data while ensuring it is not poorly extrapolated as might be the case were the means of the covariate values to be used.

\FloatBarrier

\section[Discussion]{Discussion}
\label{sec-discussion}

The \pkg{bayesics} R package provides a unified framework for performing and interpreting results from Bayesian analyses.  It focuses on the most commonly used approaches (hence the package name's homophone with ``basics''), and aims to provide the most important Bayesian inferential quantities, namely point estimates, interval estimates, probability of direction, probability that an estimand falls in the ROPE, and when applicable Bayes factors.

\pkg{bayesics} aims to eliminate the need for algorithmic knowledge so that the user can focus on inference.  This is achieved by using closed-form solutions when possible, and when posterior sampling is required automatically selecting the number of samples required for accurate Monte Carlo estimates.  This is a concern that the authors feel is widely ignored or underestimated by practitioners, as algorithmic quantities measuring MCMC convergence and accuracy of point estimates provided in alternative software may mislead practitioners into a false sense of accuracy in interval estimates.  

While \textit{algorithmic} assessment is not applicable to \pkg{bayesics}, \textit{model} assessment is still essential.  \pkg{bayesics} attempts to provide easy access to diagnostic plots, and should model assumptions not hold, \pkg{bayesics} provides several non-parametric or semi-parametric alternatives.

It is the authors' hope that \pkg{bayesics} will serve as a useful and accurate software implementation of highly used Bayesian analyses.  However, we anticipate that there will be countless bugs to correct and beneficial features not yet implemented, and we humbly ask the statistical community to report such issues at \url{https://github.com/dksewell/bayesics/issues}.

\bibliography{refs}

\end{document}